# Effective Pre-Silicon Verification of Processor Cores by Breaking the Bounds of Symbolic Quick Error Detection

Karthik Ganesan[1], Florian Lonsing[1], Srinivasa Shashank Nuthakki[1], Eshan Singh[1],
Mohammad Rahmani Fadiheh[2], Wolfgang Kunz[2], Dominik Stoffel[2], Clark Barrett[1],
and Subhasish Mitra[1]

[1]Stanford University, USA

[2]TU Kaiserslautern, Germany

**Abstract.** We present a novel approach to pre-silicon verification of processor designs. The purpose of pre-silicon verification is to find logic bugs in a design at an early stage and thus avoid time- and cost-intensive post-silicon debugging. Our approach relies on symbolic quick error detection (Symbolic QED, or SQED). SQED is targeted at finding logic bugs in a symbolic representation of a design by combining bounded model checking (BMC) with QED tests. QED tests are powerful in generating short sequences of instructions (traces) that trigger bugs. We extend an existing SQED approach with symbolic starting states. This way, we enable the BMC tool to select starting states arbitrarily when generating a trace. To avoid false positives, (e.g., traces starting in unreachable states that may not behave in accordance with the processor instruction-set architecture), we define constraints to restrict the set of possible starting states. We demonstrate that these constraints, together with reasonable assumptions about the system behavior, allow us to avoid false positives. Using our approach, we discovered previously unknown bugs in open-source RISC-V processor cores that existing methods cannot detect. Moreover, our novel approach outperforms existing ones in the detection of bugs having long traces and in the detection of hardware Trojans, i.e., unauthorized modifications of a design.

## 1  Introduction

Pre-silicon verification requires major effort in a typical hardware design flow [Foster 15]. We consider pre-silicon verification of single processor cores, which are critical components of any System-on-Chip (SoC). Generally, pre-silicon verification mainly targets logic design errors (*logic bugs*). However, it is also crucial to detect Hardware Trojans (*HTs*) [King 08], which are unauthorized modifications of an integrated circuit (*IC*) that result in incorrect functionality and/or the exposure of sensitive data [Karri 10]. While previous research on HTs focused on attacks implemented during fabrication, there is growing concern about HTs being inserted in third-party Intellectual Property (*IP*) cores by malicious sources [Zhang 11]. This makes HT detection challenging.

Similar to logic bugs, HTs can affect functionality, i.e., such an HT can cause an error that can ultimately create a change in the software-visible state of a system defined by the register or memory state. The objective of HT detection is to detect these changes, which encompass many catastrophic attacks on processor cores [King 08].

Symbolic quick error detection (Symbolic QED, or SQED) [Singh 18] is a new pre-silicon verification technique based on QED tests [Lin 14]. QED tests are powerful in generating short sequences of instructions (traces) that trigger bugs. SQED is targeted at finding logic bugs in a symbolic representation of a design by combining bounded model checking (BMC) [Clarke 01] with QED tests. As such, it is an automatic bug detection and localization technique that is effective in practice. SQED was recently applied to several industrial microcontroller cores used in commercial automotive products [Singh 19]. It was able to detect all recorded logic bugs in the designs, while enabling an 8-60X (depending on the design) reduction in verification effort compared to the standard industrial verification flow.

SQED analyzes a design symbolically, but it requires a *concrete* starting state, e.g., given explicitly as a bitvector of 0s and 1s. As a consequence, it is difficult in general to find bugs or HTs using SQED that require very long activation sequences (i.e., many instructions are required to activate such bugs). This is due to the fact that SQED must rely on deep BMC runs that unroll the circuit far enough to include all the activation instructions necessary to trigger the bug. However, deep unrollings may be prohibitive in practice. In a related study [Singh 18], it was shown that BMC could unroll a large, multicore SoC up to around 30 clock cycles, within 24 hours of verification time. Moreover, while being highly effective for detecting logic bugs, SQED can be insufficient for detecting HTs, as the following example [Zhang 14] shows.

*Motivating Example.* Consider the following HT that is difficult to find using existing HT detection techniques: *The HT changes opcodes of the next several decoded instructions after it sees a specific sequence of 256 instructions.* This HT could inject an instruction sequence to bypass physical memory protection and run a privileged instruction. Such privilege escalation attacks [King 08] can be catastrophic.

Because the above HT requires a long sequence of instructions and hence many clock cycles for activation, SQED, like other BMC-based methods [Rajendran 15, Reece 16], fails to detect the HT unless the selected starting state for BMC quickly transitions to a state where the HT activates. Stumbling upon such a close state by starting at a concrete state, e.g., obtained from simulation or a reset state is highly unlikely to succeed as the HT could be designed with an arbitrary activation sequence not known *a priori*.

To overcome the major challenge of deep unrollings in BMC, we extend SQED by allowing the integrated BMC tool to start from a *symbolic state* rather than a concrete one. This way, the BMC tool can choose one out of many possible starting states.

However, it is well-known that starting BMC from an unrestricted symbolic state risks generating spurious counter-examples (false positives). Existing BMC-based pre-silicon verification techniques, including SQED, face the challenge of false positives when used in conjunction with symbolic starting states. If the BMC tool selects a starting state that is not reachable from the set of all reset states of the system via a sequence of instructions, then a false positive might occur. For example, assume that each word in a memory is protected with a single even parity bit and assume that a BMC tool is asked to check the following property: *for any sequence of reads and writes to the memory, the parity bits remain consistent with the data.* If the starting state of the design is not constrained, then the BMC tool can initialize the memory to contain an all-zero word with a '1' for the parity bit, issue an instruction that reads from this location, and report this false positive. Traditional methods rely on verification engineers to (manually) create constraints to rule out such false positives, which can be time-consuming for practical designs having many complex properties.

As our main contribution, we present a novel variant of SQED with symbolic starting states that overcomes the associated challenges and avoids false positives in the context of detecting bugs that are detectable by QED tests. In our approach, the BMC tool potentially starts from states that could otherwise be reached only by deep BMC runs, which are prohibitive for existing formal tools. As a result, our approach allows us to detect logic bugs and HTs that require long activation sequences. This is made possible by starting from a state that is close to the bug and by applying a relatively short sequence of instructions to trigger the bug.

Importantly, SQED does not target single-instruction bugs (i.e., bugs such that a single instruction on a specific set of inputs always produces an incorrect result). There are many other known verification techniques that are highly effective at detecting such bugs, from both research literature [Reid 16] and in industry [Singh 19].

In our approach, we avoid false positives by defining certain *QED constraints*. These constraints restrict the set of possible starting states and are sufficient to ensure that false positives are avoided when using SQED with symbolic starting states to check single processor cores. Further, to start BMC from a restricted set of states, we introduce *QED recorders,* which record a small subset of internal signals within the processor to signal the point of time to the BMC tool when the QED constraints are satisfied. The QED recorders are integrated in the model of the system-under-test (given in some hardware description language), and hence are used for pre-silicon verification only. In particular, they do not incur area overhead in the final design.

Our work improves on a previous technique for SQED with symbolic starting states, called $S^2$QED [Fadiheh 18]. That approach differs from ours in the types of processors that the techniques apply to, the types of HTs that can be detected, and the way symbolic initial states are implemented. We will comment on these differences and compare the two approaches empirically.

Based on experimental results, we make the following observations:

1) We automatically, correctly, and quickly (~1 minute) detect several previously unknown (real) logic bugs in open-source out-of-order (OoO) superscalar [Ridecore1], and in-order scalar [Vscale] cores.
2) We automatically, correctly, and quickly (within 25 seconds for in-order, 18 minutes for OoO) detect 100% of (117 in-order, 120 OoO) simulated logic bugs, representing a wide variety of difficult bug scenarios (Appendix A) from commercial designs. SQED with a concrete starting state detects only 33% (in-order) and 5% (OoO).
3) We automatically, correctly, and quickly (within 5 minutes for an in-order core; 2 hours for an OoO superscalar core) detected 100% of (156 in-order, 195 OoO) simulated HTs, representing a wide variety of scenarios (Appendix A) in HT research literature. SQED with a concrete starting state detects 15% (in-order) and 9% (OoO).
4) We automatically, correctly, and quickly (within 2.5 hours) detected 97.9% of an extremal bug family (randomly-generated pre-condition-based bugs which require ~100,000 activation instructions taken from random test programs) in an OoO superscalar core. In contrast, SQED with a concrete starting state detected 0%.

The following are some of the important features of our technique:

1) It is highly effective for detecting both logic bugs and HTs (despite long activation sequences) during pre-silicon verification of in-order and OoO superscalar cores, as demonstrated by our results.
2) It does not require the verification engineer to manually craft design-specific assertions/constraints to detect logic bugs or HTs.
3) No false positives occurred, as demonstrated by our results.
4) It does not require a golden model or simulation data of the design-under-test for detection of logic bugs and/or HTs.
5) Its effectiveness does not depend on the way HTs are designed, i.e., our method is HT-design agnostic.

The rest of this paper is organized as follows. Sec. 2 provides background on earlier QED works. Sec. 3 describes Symbolic QED with symbolic starting states. Results are presented in Sec. 4, followed by related work and conclusions in Sec. 5. and 6.

## 2 Background

In the following, we present the basics of SQED and related terminology.

### 2.1 QED and the EDDI-V Transformation

Quick error detection (QED) is a testing technique that takes existing system validation tests (i.e., sequences of instructions) and automatically transforms them into a set of new tests using various QED transformations [Lin 14]. QED aims at detecting bugs which require activation sequences. To this end, several QED transformations may be applied. Among them, the Error Detection using Duplicated instructions for Validation (EDDI-V) transformation [Lin 14] is the focus of our work. It targets bugs inside processor cores by checking the results of *original* instructions against the results of *duplicate* instructions. First, the registers and memory space are divided into two halves, one for the original and one for the duplicated instructions. Next, corresponding registers and memory locations for the original and the duplicated instructions are initialized to hold the same values. This is called a *QED-consistent* system state. Then, for every load, store, arithmetic, logical, shift, or move instruction in the original test, EDDI-V creates a corresponding duplicate instruction that performs the same operation, but uses the registers and memory reserved for duplicate ones. The duplicated instructions execute in the same relative order as the original ones, but may be interleaved. The EDDI-V transformation also inserts periodic check instructions that compare the results of the original instructions against those of the duplicated ones. A mismatch in any check indicates an error and the respective instruction sequence constitutes a trace.

### 2.2 Symbolic QED

Symbolic QED (SQED) [Singh 18] combines QED transformations with *Bounded Model Checking* (BMC) [Biere 99, Clarke 01]. SQED creates a BMC problem to search through *all possible* EDDI-V tests within a bounded number of clock cycles. Thereby, the BMC tool selects which symbolic instructions to consider for execution. It searches for counterexamples to properties of form: Ra==Ra'. Here, Ra is an original register, and Ra' is the corresponding duplicate register in an EDDI-V test. To ensure that counterexamples are *QED-compatible* (we consider only EDDI-V tests), the original instructions must be valid instructions from the instruction set of the design-under-test, and the instruction sequence must be an EDDI-V test.

To implement EDDI-V tests in SQED, a *QED module* is integrated in the model of the design-under-test. The QED module automatically transforms a given input sequence of original instructions into a QED-compatible instruction sequence (e.g., as in Fig. 1). The QED module requires that the input sequence consists of valid instructions that read from or write to only the original registers and memory. These conditions are part of the BMC check. After execution of original and duplicate instructions, the QED module asserts a signal (*qed_ready*) indicating that all original and corresponding duplicate registers should contain the same values under the assumption of a bug-free system. Once signal *qed_ready* has been asserted, the BMC tool checks the property:

$$qed\_ready \rightarrow \bigwedge_{a \in \{0..\frac{N}{2}-1\}} \mathbf{Ra} == \mathbf{Ra'},$$

where N is the number of registers defined by the instruction set architecture. Here (for $\mathbf{a} \in \{0..N/2-1\}$), $\mathbf{Ra}$ and $\mathbf{Ra'}$ correspond to original and duplicate registers.

```
                          R1  = R1  + 5
                          R3  = R1  * R2
        R1 = R1 + 5       R17 = R17 + 5
        R3 = R1 * R2      R19 = R17 * R18
            (a)                  (b)
```

**Fig. 1.** Example of QED transformation by the QED module. (a) A sequence of original instructions, and (b) transformed instructions executed by the processor. R17, R18, and R19 are the duplicate registers of R1, R2, and R3, respectively.

The starting state for the BMC run must also be a *QED-consistent* state, in which the value stored in each original register or memory location matches the corresponding duplicate register or memory location. This is to prevent false counter-examples from being generated. One way to obtain such a state is to run an EDDI-V test in simulation and stop immediately after QED checks have compared all register and memory values.

Symbolic QED can detect HTs if it finds an EDDI-V test for which the HT affects original registers and duplicate registers differently. For example, assume a HT is inserted that is activated when a 128-bit counter reaches its maximum value and which changes an in-flight instruction to a NOP (cf., activation criteria A.2.a.2 ($X_I$=128) of Appendix A and effect A.2.b.1 of Appendix A). If the counter is initialized to $2^{128}$-1, and the register file is initialized at a QED-consistent state, SQED can detect the HT using the EDDI-V test {ADDI R1, R1, 2; ADDI R17, R17, 2; CHECK R1==R17}. However, as the existence of the counter is unknown it is impossible to pick the concrete starting state *a priori*. Our approach of SQED with symbolic starting states detailed in Sec. 3 below automatically detects this HT by starting the design at state {R1=0, R17=0, HT counter initialized to 1 cycle before activation}, and running the EDDI-V test.

## 2.3 S²QED

S²QED [Fadiheh 18] is another technique for pre-silicon verification which extends SQED by incorporating symbolic starting states. Like our approach, S²QED focuses only on EDDI-V tests. S²QED instantiates two copies of the CPU, called CPU-1 and CPU-2. An arbitrary one-to-one mapping is then determined between the registers of the two CPU instances. A new notion of *QED consistency* is defined when all values in the general- and special-purpose registers of CPU-1 match the values in the mapped registers of CPU-2.

At the start of verification, CPU-1 fetches an instruction called the instruction under verification (IUV) while CPU-2 fetches the corresponding QED duplicate instruction. The duplicate instruction has each register of CPU-1 replaced by the mapped register in CPU-2. Then, CPU 1 is constrained to fetch NOPs until the IUV commits, while CPU-2 can fetch arbitrary valid instructions (treated as symbolic instructions by the BMC tool). S²QED then attempts to prove that at commit time for the IUV, the registers in the two CPUs will always remain QED consistent. Thus, S²QED is able to prove that the model of the processor design is free of bugs of a specific class.

The following example shows a bug that can only be caught by our approach but not by S²QED. *When all registers in the register file contain the value -1, the next register write is corrupted.* This bug escapes S²QED, because it affects the original and duplicate instructions equivalently. In contrast, our approach creates scenarios where a mismatch between an original and duplicate instruction occurs. We provide a related experimental comparison with more examples in Sec. 4.4.

Our approach also differs from S²QED in that it is especially suited for a broader class of processor designs, including OoO superscalar processors. This capability is enabled by the QED constraints we define (Sec. 3.2), together with QED recorders (Sec. 3.3) and a new QED module (Sec. 3.1). In contrast to our approach, S²QED is applicable to processors with Out-of-Order writeback. This is possible due to additional constraints that restrict the state of the instruction pipeline. Such constraints can also be integrated in our framework. Further, S²QED does not require a QED module or QED recorders. However, it requires duplication of the CPU in the model of the design-under-test (only during pre-silicon verification), whereas our approach requires only a single CPU.

## 3 Extending Symbolic QED with Symbolic Starting States

In the following, we present our extension of Symbolic QED [Singh 18] with symbolic starting states. First, we describe the design of a new QED module that is integrated in the model of the design-under-test during pre-silicon verification. The QED module controls the QED tests carried out by the BMC tool. Our new QED module improves upon related work [Singh 18] as it enables the detection of some of the logic bugs in our experimental study (Sec. 4). To avoid false positives, we define a set of constraints (*QED constraints*) on the symbolic starting state (Sec. 3.2). To implement the QED constraints, we introduce *QED recorders*. These recorders are additional hardware modules added to the model of the design (only in pre-silicon verification—not for the final design) that record a small subset of internal logic values of the processor core to ensure that the QED constraints are satisfied when a QED test begins in the BMC tool (Sec. 3.3). Fig. 2 contrasts Symbolic QED with/without symbolic starting states.

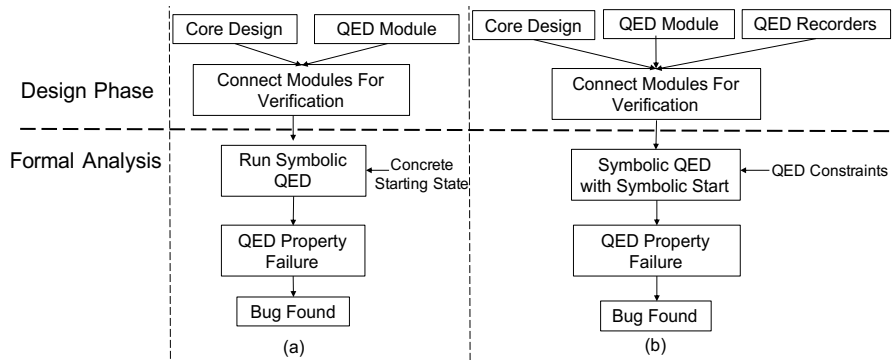

**Fig. 2.** (a) Symbolic QED inputs and steps without symbolic starting state, and (b) Symbolic QED inputs and steps with symbolic starting state.

### 3.1 New QED Module for Single Processor Cores

Pseudocode for the new QED module is given in Fig. 3(a). Inputs are: 1) *enable*: disables the QED module if set to 0; 2) *next_instruction*: next instruction to be executed; 3) *fetch_next*: high when the core is ready to receive an instruction, i.e., the fetch stage is not stalled; 4) *original*: tells the core to execute an original (if high) or duplicate (if low) instruction. Outputs are: 1) *instruction_valid*: indicates whether the output instruction is valid; and 2) *instruction_out*: instruction to be executed. The QED module has internal variables: 1) *queue*: a queue data structure that stores previous original instructions that have not yet been executed in the duplicate subsequence; 2) *head_instruction*: the previous head of the queue; 3) *insert_valid*: true when the QED module can execute an original instruction; 4) *delete_valid*: true when the QED module can execute a duplicate instruction; 5) *duplicate_instruction* next instruction in the duplicate subsequence to be executed (when *original* is 0).

In SQED, QED checks are carried out in QED-consistent system states. Such states are determined by the *qed_ready* signal of the QED module. Pseudocode for determining this signal is given in Fig. 3(b). To avoid trivial false positives, QED checks occur when an equal number of commits (writes) have been made to original registers and duplicate registers. This is accomplished by keeping track of the number of original and duplicate commits to the register set, as shown in Fig. 3(b). For simplicity, in Fig. 3(b), we assume that at most one instruction commits per cycle. For superscalar processors that can commit multiple instructions in the same cycle, we track all corresponding pairs of *write_valid* (tells whether the input data is valid) and *write_address* (the address for the data to be written) signals, keep a separate *is_original* signal (identifies if an address corresponds to an original or duplicate location) for each instruction, and allow the original and duplicate counters to be incremented multiple times if needed.

The QED module of [Singh 18] requires that all original instructions complete, a waiting period occurs for the pipeline to be flushed, and duplicate instructions execute, before the *qed_ready* signal is asserted. In contrast to that, our new QED module allows arbitrary interleaving of original and duplicate instruction subsequences without a waiting period. This is made possible by giving the BMC tool control over the *original* input of Fig. 3(a). The QED ready enable logic (Fig. 3(b)) can be further enhanced as follows:

1. The current QED ready enable logic is only applicable to single processor cores, since a multi-core system would require modification of the *qed_ready* logic to consider the original and duplicate commits across all cores. This can be challenging in situations where multiple cores operate with a shared address space. For simplicity, we do not consider this situation in this paper.

2. For some processors, e.g., superscalar processors with explicit register renaming (MIPS 10000 [MIPS 96] and ARM's Cortex-A15 [ARM]), the designation of original or duplicate instruction cannot be made solely on where they write (unlike in Fig. 3(b)). This issue can be corrected by including the current state of the register mapping table as an input to the function is_write_to_original_space. Each time a QED check happens, the same mapping table must be used to map logical to physical addresses before comparing "original" and "duplicate" values. The RISC-V cores used in our experimental evaluation, however, do not have this issue.

### 3.2 QED Constraints

We first define some terminology used in the constraint definitions: i) *Symbolic In-Flight (SIF) "instructions"*: symbols (i.e., state bits), part of the symbolic starting state (which will be assigned 0s and 1s by the BMC tool), corresponding to (microarchitectural) flip-flops within the pipeline that hold instructions during normal operation of the core[1]; ii) $T_C$: the point in time when all SIF instructions commit (i.e., write to architectural state) - it is determined by the BMC tool; iii) *Symbolic QED instructions*: symbols which represent the instructions that form the bug trace (which is part of the counter-example, along with the starting state that BMC assigns) generated by the BMC tool after $T_C$; and iv) *Symbolic QED operand data:* symbols representing the operand[2] data of dispatched Symbolic QED instructions (dispatched before $T_C$).

```
INPUT: enable, next_instruction, fetch_next, original
OUTPUT: instruction_out, instruction_valid
// initialization
queue ← 0;  head_instruction ← 0;
// end initialization
insert_valid ← fetch_next & original & ~queue.is_full();
delete_valid ← fetch_next & ~original & ~queue.is_empty();
instruction_valid ← insert_valid | delete_valid;
if insert_valid then
  queue.push(next_instruction);   // store this instruction in queue
else if delete_valid then
  head_instruction ← queue.pop(); // remove instruction at the head from queue
end if
duplicate_instruction ← create_duplicated_version(head_instruction);
instruction_out ← (enable & ~original) ? duplicate_instruction : next_instruction;
```
(a)

```
INPUT: write_valid, write_address
OUTPUT: qed_ready
// initialization
qed_ready ← false;  count_original ← 0;  count_duplicate ← 0;
// end initialization
is_original ( is_write_to_original_space(write_address);
if write_valid then
  if is_original then
    count_original++;     // increment number of original instructions committed
  else
    count_duplicate++;    // increment number of duplicate instructions committed
  end if
end if
qed_ready ( (count_original == count_duplicate) ? true : false;
```
(b)

**Fig. 3.** Pseudocode for (a) QED module, and (b) QED ready enable logic

Fig. 4 illustrates these definitions for a 3-stage in-order pipeline. When the formal analysis begins, there are up to 3 SIF instructions, and all commit by time $T_C$. The first Symbolic QED instruction (corresponding to *R1=R1+5*) is fetched into the pipeline, and its Symbolic QED operand data is available after the Dispatch stage.

Now, the QED constraints are given as follows (Appendix C further details how each constraint is enforced):

**Constraint C-1:** At $T_C$, all SIF instructions have committed (i.e., no SIF instruction can write to the architectural state after $T_C$), while all Symbolic QED instructions commit after $T_C$.

**Constraint C-2:** At $T_C$, the architectural state (program-visible registers and memory) is QED-consistent (Sec. 2.2), and nothing but Symbolic QED instructions can write to architectural state after $T_C$ (e.g., any test mode such as scan that can directly write to flip-flops must be disabled).

**Constraint C-3:** All the operand data for each Symbolic QED instruction I, must satisfy one of the following properties:

i) if the operand data is available (i.e., I has already read data for this operand) then this data must match the corresponding register/memory location (i.e., source operand location) data at $T_C$.

ii)[3] if the operand data is not available at $T_C$, then I is waiting for the result of an earlier Symbolic QED instruction for this operand data.

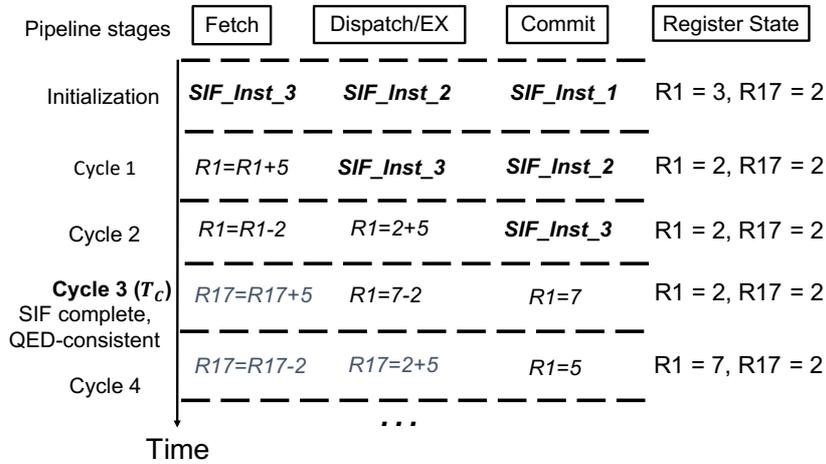

**Fig. 4.** Timing diagram for a three-stage in-order pipeline satisfying all QED constraints. SIF instructions commit by Tc, before all QED instructions.

The QED constraints form a sufficient condition to ensure no false positives, given that bug-free designs satisfy two assumptions after $T_C$:

1. If a Symbolic QED instruction is executed twice on the same data, it results in the same value being stored to architectural state, e.g., Rx=1+2 and Ry=1+2 always result in the same values stored to registers Rx, Ry.
2. If a Symbolic QED instruction has a read-after-write dependency with earlier instructions, it uses the most recent value of the data in its computation. For example, in the program {R1=5; R2=R1+2; R3=R2-2}, the second instruction uses value '5' for R1, and the third instruction uses value '7' for R2.

We observed empirically (see Sec. 4), that at least one of the assumptions was violated in traces obtained from BMC. The two assumptions are further detailed in Appendix B.

### 3.3 Symbolic QED Recorders

**Recorder for $T_C$.** As $T_C$ depends on the starting state chosen by the BMC tool, it cannot be statically determined before the formal analysis begins. A recorder is used to give this information to the tool dynamically. For an in-order core, $T_C$ can be determined by simply tracking the progress of the first Symbolic QED instruction (the first symbolic instruction the BMC tool creates as part of the bug trace) until it reaches the commit stage (write-back stage) of the pipeline. At this time, all SIF instructions must have committed, as the pipeline is occupied by Symbolic QED instructions.

Specifics of the $T_C$ recorder for a 5-stage, single-fetch, in-order pipeline is given in Fig. 5. Inputs are ready signals for all stages that precede the commit stage (e.g., *fetch_ready* is high when the fetch stage is ready to receive an instruction). The output *SIF_complete* is true when the first Symbolic QED instruction goes through all pipeline stages and reaches the commit stage. The output *mode* keeps track of progress made so far by the Symbolic QED instruction (we later make use of this output in the Symbolic QED operand recorder). This $T_C$ recorder for a 5-stage pipeline can be easily modified to support in-order pipelines with a different number of stages.

For an OoO core, the $T_C$ recorder is even simpler, and utilizes the reorder buffer (ROB). The idea is to mark the entry allocated in the ROB for the first Symbolic QED instruction. After this, *SIF_complete* is assigned true when the ROB head pointer reaches the marked instruction.

For cores with no ROB, but OoO *commit* (e.g., [Aquarius]), an additional constraint is required (see Sec. 2.3).

```
INPUT: fetch_ready, dispatch_ready, exec_ready, mem_ready
OUTPUT: SIF_complete, mode
// initialization
mode ← S₀;   SIF_complete ← false;
// end initialization
if (mode == S₀) && (fetch_ready) then
  mode ← S₁;   SIF_complete ← false;   // instruction completes fetch stage
end if
if (mode == S₁) && (dispatch_ready) then
  mode ← S₂;   SIF_complete ← false;   // instruction completes decode stage
end if
if (mode == S₂) && (exec_ready) then
  mode ← S₃;   SIF_complete ← false;   // instruction completes execute stage
end if
if (mode == S₃) && (mem_ready) then
  mode ← S₄;   SIF_complete ← true;   // instruction completes memory stage
end if
```

**Fig. 5.** Pseudocode for $T_C$ recorder.

**Recorder for Symbolic QED Operands.** Like $T_C$, Symbolic QED operands also depend on the starting state. The Symbolic QED operand recorder stores information for both register and memory operands. Specifics of the Symbolic QED operand recorder for a 5-stage, single-fetch, in-order pipeline is given in Fig. 6. Inputs are: 1) *\*_addr*, which gives register/memory address of the corresponding operand; 2) *\*_data*, which gives operand data; 3) *\*_valid*, which is high when *\*_addr* is valid and *\*_data* is valid; 4) *mode*, which gives the state of the $T_C$ recorder (Fig. 5). Output *\*_buffer* stores all Symbolic QED operands and their values (buffer depth is determined by the maximum number of instructions in flight at a given time). We only store the information for Symbolic QED instruction operands in buffers i.e., we do not store operand information of any SIF instruction. This is enforced by checking the $T_C$ recorder state, i.e., *mode* (we do not add entries to *_buffer until all SIF instructions pass through dispatch stage). In Fig. 6, we assume that each instruction requires at most two register values and one memory value, but this idea is trivially extended to more source operands.

For an OoO core, Fig. 6 is extended to include Symbolic QED operands that are waiting on results of earlier Symbolic QED instructions. For each waiting operand, we store the instruction tag (ROB entry number) of the instruction it is waiting for. This information is used to specify Constraint C-3 for an OoO core (details in Appendix C).

```
INPUT: src1_addr, src1_data, src1_valid, src2_addr, src2_data, src2_valid,
mem_addr, mem_data, mem_valid, mode
OUTPUT: src1_buffer, src2_buffer, mem_buffer
// initialization
src1_buffer ← empty_buffer;  src2_buffer ← empty_buffer;
mem_buffer ← empty_buffer;
// end initialization
if (mode != S0) && (mode != S1) && (src1_valid || src2_valid) then
  if (src1_valid) then
    src1_buffer.add_entry(src1_addr, src1_data);
  end if
  if (src2_valid) then
    src2_buffer.add_entry(src2_addr, src2_data);
  end if
end if
if ((mode == S3) || (mode == S4)) && (mem_valid) then
  mem_buffer.add_entry(mem_addr, mem_data);
end if
```

**Fig. 6.** Pseudocode for Symbolic QED operand recorder.

## 4  Results

We demonstrate the effectiveness of our new technique on two open-source RISC-V processor cores: i) V-scale [Vscale], an in-order core targeting embedded applications; and ii) RIDECORE [Ridecore1], an OoO superscalar core (2-way pipeline, 64 maximum instructions in-flight, 2 ALUs, 1 multiplier, 1 load/store unit) for high performance applications. For BMC, we used the Questa Formal tool (version 10.5c) from Mentor Graphics on an AMD Opteron 6438 with 128GB of RAM. For each core, we instrumented a new QED module (Sec. 3.1) and the QED recorders and QED constraints (Sec. 3).

### 4.1  Previously Unknown Bugs

We found three previously unknown logic bugs in the multiplier reservation station (RS-m) of the RIDECORE design, all of which were confirmed by RIDECORE designers [Ridecore2], see Table 1. Importantly, these bugs were detected due to our new QED module (Sec. 3.1). The QED module of [Singh 18] is not capable of detecting these bugs. They require original and duplicate multiply instructions to execute on subsequent clock cycles, while the QED module of [Singh 18] requires a waiting period between original and duplicate instructions. Our new QED module improves upon [Singh 18] by allowing arbitrary interleaving of original and duplicate instructions.

**Table 1**. New bugs in RIDECORE. Symbolic QED runtimes.

| Bug Activation | Bug Effect | Runtime (symbolic starting state) | Runtime (power-on reset starting state) |
|---|---|---|---|
| All but one (buggy entry) RS-m entries occupied; MULH[4] instruction assigned to vacant entry. | First source operand of MULH instruction corrupted. | 25 minutes | 63 seconds |
| Same as above. | Second source operand of MULH instruction corrupted. | 61 minutes | 69 seconds |
| Same as above, but MULHU[5] instr. assigned to vacant entry. | Result of MULHU instruction corrupted. | 64 minutes | 93 seconds |

We also found two bugs in Vscale (Table 2), by running Symbolic QED starting at a concrete, power-on reset state in less than 40 seconds (also confirmed by designers). These bugs are due to errors in the Vscale implementation of the RISC-V privileged ISA [RISCVP], within specific Control Status Registers (CSRs). Importantly, Vscale does not implement shadows for CSRs. To circumvent this, Symbolic QED's EDDI-V transformation (Sec. 2.1) duplicates instructions using data memory for each CSR. The first bug occurs because of incorrect design of the MIP register interrupt bit logic. After this bug was fixed, a second bug was found in the MSTATUS register.

**Table 2**. Confirmed bugs in VSCALE. Runtimes are for Symbolic QED with concrete, power-on reset starting state.

| Bug Activation | Bug Effect | Runtime |
|---|---|---|
| The value '1' is written to specific bit positions in the machine-interrupt CSR MIP. | MTIMECMP register corrupted; Causes repeated interrupts. | 2 seconds |
| Any value with lower two bits '01' or '10' written to the machine-level CSR MSTATUS. | Design enters unspecified privilege level; MEPC register corrupted; | 33 seconds |

---

[1] The formal tool is free to choose any values for symbols (state bits) associated with in-flight instructions, including those that are not consistent with the logic driving those symbols. Thus, values chosen for symbols in a SIF instruction need not constitute a valid instruction.
[2] Operands may come from either registers or memory locations. For register (memory) operands, the dispatch stage is the register read (memory read) stage.
[3] This condition is required for OoO cores, where there is a possibility that the Symbolic QED operand may wait on a SIF instruction instead of a Symbolic QED instruction.
[4] MULH is a signed multiply instruction selecting the upper half of the multiplier result.
[5] MULHU is an unsigned multiply instruction selecting the upper half of the multiplier result.
[6] This program comes packaged with the RIDECORE design by the designers as a part of a testbench. It was used only for "extremal" bug creation – not for verification or bug detection.

### 4.2 "Long" Logic Bugs and HT Scenarios

We simulated 120 (117) logic bug types using RIDECORE (Vscale). These are mostly "longer" (up to 256 consecutive activation instructions) versions of "difficult" logic bug scenarios (Appendix A) that occurred in various commercial SoCs [Singh 18]. We also simulated 195 (156) difficult HT scenarios (Appendix A) from research literature using RIDECORE (Vscale). Results are in Table 3.

**Table 3.** "Long" logic bugs and HTs. We report [min, average, max].

|  |  | "Long" Bugs | HTs |
|---|---|---|---|
| **Vscale** | Total count injected | 117 | 156 |
|  | **Symbolic QED with symbolic starting state** | | |
|  | Coverage | 100% | 100% |
|  | Bug trace length (instructions) | [2, 2, 3] | [2, 2, 3] |
|  | Bug trace length (clock cycles) | [5, 5, 6] | [5, 5, 6] |
|  | BMC runtime (seconds) | [2, 4, 25] | [2, 11, 313] |
|  | **Symbolic QED with concrete starting state** | | |
|  | Coverage | 33% | 15.3% |
| **RIDECORE** | Total count injected | 120 | 195 |
|  | **Symbolic QED with symbolic starting state** | | |
|  | Coverage | 100% | 100% |
|  | Bug trace length (instructions) | [4, 4, 4] | [4, 4, 4] |
|  | Bug trace length (clock cycles) | [8, 8, 8] | [8, 8, 8] |
|  | BMC runtime (minutes) | [7, 13, 18] | [7, 20, 121] |
|  | **Symbolic QED with concrete starting state** | | |
|  | Coverage | 5% | 8.7% |

**Observation 1:** SQED with symbolic starting states correctly and automatically found all "long" logic bugs, in less than 30 mins, with no false positives. It found bugs that traditional BMC methods fail to detect (including SQED). SQED with concrete starting state detected only 5% (33%) of these bugs in RIDECORE (Vscale).

**Observation 2:** SQED with symbolic starting states correctly and automatically found all injected HTs (including those designed to evade state-of-the-art HT detection techniques), in less than 2.5 hours, without requiring design-specific assertions or debug of false positives. SQED with a concrete starting state detected only 9% (15%) of these HTs in RIDECORE (Vscale).

### 4.3 "Extremal" Bugs

To demonstrate the robustness of our presented technique, we inject "extremal" bugs (only triggered when the design reaches a very specific set of states) into RIDECORE (since it is OoO, superscalar, and more complex than Vscale). Our extremal bug injection methodology is as follows: i) Run Matrix Multiply (1M cycles) on the design in simulation[6] and stop the simulation at a random point in time; ii) Run a uniform random sequence of 100 ALU or Load/Store instructions; iii) Select a uniformly random subset of flip-flops from the set of all flip-flops in the design and record their logic values; and iv) Generate a bug (effect A.1.b.3 of Appendix A), injected into the design. This bug is only activated when the design reaches a state where all the selected flip flops (step iii) have a specific set of values recorded (step iii).

We present our results in Table 4. For generating such extremal bugs, we randomly chose 180 time points (step i), ranging from 26,026 to 988,159 clock cycles elapsed from program start. For each time point, we ran a random 100-instruction sequence (step ii), and then randomly selected 10 different subsets of 128 flip-flops (step iii), resulting in 1,800 total extremal bug count.

Whereas Symbolic QED with concrete starting state detected 0% of these 1,800 "extremal" bugs, Symbolic QED with symbolic starting state was able to detect 1,763 of the 1,800 bugs. For the remaining cases, the BMC tool timed out after 24 hours. A closer inspection reveals that the BMC tool was not able to unroll the design beyond 7 clock cycles (8 cycles are needed to observe these bugs). In future work, we plan to investigate ways to improve BMC tools to address such issues (following approaches such as [Ganai 04, 06]).

**Table 4.** "Extremal" logic bugs for RIDECORE. We report [min, average, max].

| Total count injected | 1,800 |
|---|---|
| **Symbolic QED with symbolic starting state** | |
| Coverage | 97.9% |
| Bug trace length (instructions) | [4, 4, 4] |
| Bug trace length (clock cycles) | [8, 8, 8] |
| BMC runtime (minutes) | [8, 33, 149] |
| **Symbolic QED with concrete starting state** | |
| Coverage | 0% |

**Observation 4:** Our new Symbolic QED with symbolic starting states correctly and automatically found 97.9% of the "extremal" logic bugs and generated a bug trace in less than 3 hours. In contrast, Symbolic QED with concrete starting state detected 0% of the extremal bugs.

### 4.4 Detection of Trojans that Evade S²QED

We constructed two families of HTs that are able to evade S²QED [Fadiheh 18], but are detected by our new technique, and injected 8 examples into Vscale (S²QED is not applicable to RIDECORE, since it does not allow OoO operations). The first set of such HTs are triggered when all general-purpose registers hold the same value (discussed in Sec. 2.3). S²QED begins in a state where general-purpose registers of CPU-1 hold the same values as corresponding mapped registers in CPU-2. Regardless of which mapping is chosen between registers of CPU-1 and CPU-2, the HT will activate in both CPUs and a discrepancy cannot be detected. The second set of HTs are triggered by a specific state of special-purpose registers. Because S²QED also initializes both CPUs in states where special-purpose registers have the same value, these HTs also effect both CPUs equivalently. However, many HTs are still expected to be caught by S²QED, and we plan on performing a more comprehensive comparison between the techniques in future work. Details on these HT constructions are given in Table 5.

**Table 5.** HT constructions in Vscale and coverage comparison with S²QED.

| HT Activation Criteria | HT Effect(s) | Count | (S²QED) Coverage | (New Technique) Coverage; |
|---|---|---|---|---|

16| | | | | |
|---|---|---|---|---|
| All general-purpose registers hold value '20' (Appendix A: A.2.a.3) | Any effect from A.2.b.1 – A.2.b.4 in Appendix A. | 4 | 0%; | 100%; |
| Control-status register 'mtimecmp' holds value 32'hFFFFFFFF (Appendix A: A.2.a.3) | Any effect from A.2.b.1 – A.2.b.4 in Appendix A. | 4 | 0%; | 100%; |

## 5  Related Work

Existing formal verification techniques employing BMC [Reid 16, Singh 18] have issues in detecting bugs that require a long activation sequence. Other works for processor cores use theorem proving [Bhadra 07], or try to learn invariants on the design [Thalmaier 10] to be used as constraints, but these techniques tend to be ad-hoc and require a high level of manual effort. In seminal work [Burch 94] and extensions [Berezin 98], models of processors were verified based on abstractions by uninterpreted functions with equality. That approach in general requires to provide invariants to avoid false positives. E-QED [Singh 17] is a BMC-based technique for electrical bug localization in *post-silicon* verification. Apart from that it is substantially different from our technique, e.g., as it does not rely on the duplication of instructions.

False positives are a major challenge for traditional BMC. However, the same QED constraints as used in our approach may not prevent false positives for general property checking using BMC. The following example illustrates this point. Let a processor core start at a state where the Exception Program Counter (*EPC*) (i.e., the register storing the return address for an exception) is misaligned, i.e., not aligned with any word in the instruction cache, the current PC is within an exception handling routine, and there are only NOP instructions in the pipeline. This is an unreachable state for processors with strict alignment rules (e.g., MIPS). It is reasonable to check the property that the EPC is aligned, since returning to a misaligned address can cause programs to crash. Even at time Tc, when the NOP sequence is finished, this EPC will still be misaligned, causing a false positive. With QED constraints in SQED though, we do not get such a false positive because the exception handling routine will be filled by valid QED tests. Hence any time we assert a QED check, it won't fail unless there is a bug in the design.

Existing HT detection techniques that can be applied in pre-silicon verification broadly belong to two categories: i) design analysis methods; and ii) formal methods [Xiao 16]. One class of design analysis techniques use the observation that signals associated with HTs may be mostly unused or rare. [Cakır 15, Hicks 10, Zhang 11, Zhang 13b] use simulation data along with rareness metrics (e.g., code coverage, signal correlation). [Waksman 13, Yao 15] do not need simulation data, but still trade off false-positives (i.e., spurious detection of HTs) for false-negatives (i.e., failure to detect HTs) and vice-versa, depending on the thresholds set for their rareness metrics.

Additionally, stealthy HTs have been designed [Zhang 14] to bypass such analyses. In contrast to that, our technique does not require simulation data, detects stealthy HTs given in [Salmani 13, Zhang 13a, Zhang 14], and does not produce any false positives. However, our technique is for processor cores, while the aforementioned analysis techniques are applicable for general designs.

Formal methods for finding HTs generally either use BMC [Rajendran 15, 16], SAT-based equivalence checking [Banga 10, Reece 16, Shrestha 12] or theorem proving [Guo 17, Jin 13, Love 12]. These techniques face similar challenges as BMC-based techniques for bug detection, especially those for detecting bugs and HTs that require long activation sequences, in addition to manual creation of properties.

Complementary approaches to ours include methods to detect HTs that leak sensitive data but do not produce incorrect logic values [Fern 17, Hu 16, Jin 12, Rajendran 16], and HT prevention techniques [Chakraborty 09, Dupuis 14, Samimi 16].

# 6    Conclusion

We extended Symbolic QED (SQED) to include symbolic starting states. Thereby, we overcome limitations of existing pre-silicon verification techniques for detecting logic bugs and hardware trojans (HTs) that require long activation sequences. This is achieved by the unique combination of SQED and novel QED constraints. These constraints allow to avoid false positives as they restrict the starting states. Our results on multiple open-source RISC-V processor cores demonstrate the effectiveness and practicality of our approach: (i) detection of previously unknown logic bugs within minutes; (ii) detection of 100% of hundreds of long logic bugs and HTs (SQED with a concrete starting state detects, at best, 33%); (iii) detection of 97.9% of extremal logic bugs (SQED with a concrete starting state detects 0%). Future research directions include: i) extending our approach to detect bugs and HTs in other SoC components beyond processor cores, such as uncore components and accelerators; ii) handling other QED transformations beyond EDDI-V, e.g., CFTSS-V and CFCSS-V [Singh 18]; iii) automated methods for inserting QED recorders and generating QED constraints on the symbolic starting state; iv) formally analyzing the bug detection capabilities and guarantees on avoiding false positives of our approach.

## Appendix A. Logic Bug and Hardware Trojan Scenarios

In the following tables, we give the different logic bug (harder versions of "difficult" bugs that occurred in various commercial designs [Singh 18]) and HT scenarios (from research literature) used in Table 3 of Sec. 4. Each "long" logic bug is modeled with two parts: i) activation criteria of the bug (Table A.1.a), i.e., the conditions which need to be satisfied for the bug to activate; and ii) effect of the bug once it is activated (Table A.1.b). For our experiments, we considered a whole range of values for the parameters in Table A.1, as follows, $N=Y=\{2,4,8,16,32,64,128,256\}$, $R=X=\{2,4,6,…,30\}$. This results in a total of 117 logic bugs in Vscale (Activation A.1.a.5 is not possible), and 120 logic bugs in RIDECORE.

**Table A.1.a.** Activation criteria for "long" logic bugs.

| | |
|---|---|
| Processor Core | 1. Data forwarding between pipeline stages. |
| | 2. Two specific instructions within $X$ cycles. |
| | 3. $R$ registers must each contain a specific value $V$. |
| | 4. A specific sequence of $N$ instructions must execute within $Y$ cycles. |
| | 5. A specific cache state. |

**Table A.1.b.** Bug effect from [Singh 18].

| | |
|---|---|
| Processor Core | 1. Next instruction corrupted to NOP. |
| | 2. Next instruction opcode incorrectly decoded. |

|  | 3. Next instruction register read corrupted. |
|---|---|

**Table A.2.a.** Activation criteria for HTs from [Salmani 13].

| Processor Core | 1. Specific length $N$ sequence on $M_1$ internal wires. |
|---|---|
|  | 2. $X_1$ bit counter reaching final value. |
|  | 3. Comparator on $M_2$ internal wires becomes true. |
|  | 4. $X_2$ bit rare event counter reaches a specific value. |

**Table A.2.b** HT effects

| Processor Core | 1. An in-flight instruction changed to NOP. |
|---|---|
|  | 2. Opcode of an in-flight instruction changed. |
|  | 3. Next register read corrupted. |
|  | 4. Next result of an execution unit changed. |
|  | 5. Corrupts ROB. Prematurely commits next inst. |

**Table A.2.c.** HT Design Techniques

| Methodology | HT stealthy against following techniques |
|---|---|
| [Salmani 13] | Traditional pre-silicon verification techniques. |
| [Zhang 13a] | UCI [Hicks 10]; coverage metrics [Hicks 10, Zhang 11]. |
| [Zhang 14] | [Hicks 10, Zhang 11, Waksman 13, Zhang 13b]. |

Table A.2.a gives HT activation scenarios that capture HT triggering mechanisms in the Trust-Hub benchmarks [Salmani 13]. Table A.2.b gives various effects an HT can have on the executing instructions [King 08]. Table A.2.c presents three HT implementation techniques used to inject HTs in designs [Rajendran 15]. We create stealthy HTs that are known to evade common detection techniques (e.g., HT designs from [Zhang 13a] evade detection techniques based on UCI [Hicks 10] and coverage metrics [Hicks 10, Zhang 11]). A HT scenario is formed by using one activation criteria (Table A.2.a) with one bug effect (Table A.2.b), along with an appropriate design strategy (Table A.2.c). We used a wide range of HT scenario parameters, given in Table A.2: $N=\{2,4,8,…,256\}$, $M_1=32$, $X_1=X_2=\{128,256\}$, $M_2=64$, resulting in 156 HT scenarios in Vscale (effect A.2.b.5 is not possible) and 195 HT scenarios in RIDECORE. These values make HTs harder to activate than benchmark HTs in [Salmani 13].

Bugs and HTs were injected by introducing a small state machine into the design that checks for the activation criteria, and flips bits at a desired wire in the design to achieve the effect.

## Appendix B. Assumptions on Bug-Free Designs

In this Appendix, we detail our assumptions on how any bug-free design should operate. We relate these assumptions to the QED constraints of Sec. 3, and explain how they are used to prevent false positives.

<u>Assumption-1:</u> Let $I_1$ and $I_2$ be two Symbolic QED instructions given abstractly as:
$$I_1 : d \leftarrow op\ s_1, s_2, …, s_m$$
$$I_2 : d' \leftarrow op\ s_1', s_2', …, s_m'.$$

Here op represents an operation performed on data stored in the operand locations ($s_1$, $s_1'$ etc.), and the instructions write the computed result to a destination location in the architectural state (e.g., d, d′). Let $data(s_1, I_1)$ be a notation to represent the operand $s_1$ data used by instruction $I_1$ for computing the result, (Note the actual data for the source operand may be either obtained from an architectural state, e.g., architectural register, or a micro-architectural state (e.g., by data forwarding). $data(s, I)$ abstracts away these extra details and only represents the data value used by instruction I for operand s.) and $val(T_{I_1}, d)$ represents the value in the architectural location d immediately after instruction $I_1$ commits at time $T_{I_1}$ (i.e., the value written by instruction $I_1$ to location d).

Now, Assumption-1 states: If two Symbolic QED instructions $I_1$, $I_2$ have the same op and $\forall i \in \{1, \ldots, m\}, data(s_i, I_1) = data(s_i', I_2)$, then $val(T_{I_1}, d) = val(T_{I_2}, d')$.

This Assumption simply states that when a Symbolic QED instruction executes twice on the same data, then it should result in the same outputs.

<u>Assumption-2:</u> Let $I_1$ and $I_2$ be two Symbolic QED instructions given as below:
$$I_2 : s_i \leftarrow op_2\ s_1', s_2', \ldots, s_m'$$
$$I_1 : d \leftarrow op_1\ s_1, s_2, \ldots, s_m,$$
where $i \in \{1, \ldots, m\}$ and $I_2$ is the last Symbolic QED instruction with which $I_1$ has Read-after-Write (RAW) dependency for operand $s_i$. Then:
$$\text{if } I_1, I_2 \text{ are as above}, data(s_i, I_1) = val(T_{I_2}, s_i).$$

Further, if there are no earlier Symbolic QED instructions writing to an operand $s_i$ of $I_1$, where $i \in \{1, \ldots, m\}$, then $data(s_i, I_1) = val(T_C, s_i)$.

In words, this Assumption states that when any operand of a Symbolic QED instruction has RAW dependency with any earlier Symbolic QED instruction(s), it should obtain the correct source value (which is the result of the last instruction it is dependent on). Also, when an operand of a Symbolic QED instruction does not have dependencies with earlier Symbolic QED instructions, it obtains the correct source value from the architectural state at time $T_C$, to be used in its computation.

Both Assumptions are naturally satisfied in a bug-free processor core, given it starts from a state which is reachable from one of the specified reset states. However, these Assumptions may not hold if the core starts from an arbitrary symbolic state. Nevertheless, as Constraint C-1 guarantees that all SIF instructions commit at $T_C$, we expect them to hold after $T_C$. In other words, even if the processor pipeline starts from an unreachable state, we expect the pipeline to reach a state (after $T_C$) at which the Assumptions hold. For example, if the processor core has an in-order pipeline, then we expect the state-bits associated with the in-order pipeline stages to go to a reachable state after $T_C$, as all the SIF instructions would have been completed by then and the pipeline stages are either empty or filled with Symbolic QED instructions (which are valid instructions that have propagated through the pipeline stages).

Thus, we expect the Assumptions to hold for in-order pipeline cores. For superscalar cores also, Constraint C-1 guarantees that no SIF instructions are in the processor pipeline, which is typically made of in-order stages for fetch, decode, and commit and an out-of-order execution stage having reservation stations or instruction buffers to store in-flight instructions. At $T_C$, the state bits associated with in-order pipeline stages, along with state bits of some entries in the instruction buffers will be filled with Symbolic QED instructions, taking these state bits to valid (reachable) states. Also, at $T_C$, all remaining entries in the instruction buffers should be empty entries, otherwise, it would mean that there is an uncommitted SIF instruction which would violate Constraint C-1.

Thus, the state bits associated with these entries, at $T_C$, must be of no consequence to the execution of the processor. Thus, the overall state of a superscalar core, at $T_C$, is such that it has only valid Symbolic QED instructions in-flight. Consequently, we expect the Assumptions to also hold in a typical superscalar core. We have empirically seen that the Assumptions hold in both an in-order core [vscale] and a superscalar core [Ridecore1].

## Appendix C. Specifying Constraints to the BMC Tool

In this Appendix, we describe in detail how we specify the constraints on the symbolic initial state (introduced in Sec. 3.2.) to the BMC tool.

### C.1. Specifying C-1

Constraint C-1 states that by $T_C$, all SIF instructions have committed, while no Symbolic QED instruction has committed. This condition is naturally satisfied in processors with in-order execution, which is the case for processors with in-order pipelines.

But, this is not the case with superscalar cores. This is due to instruction indirection, i.e., renaming of instructions using respective ROB entries, which happens to support out-of-order execution inside a superscalar core. As we are starting with a symbolic state, ROB entry locations for SIF instructions may be chosen by the BMC tool such that they commit after Symbolic QED instructions commit, thereby violating C-1. So, in a superscalar core, specifying C-1 to the BMC tool involves properly constraining the ROB entries for SIF instructions to avoid the issue.

### C.2. Specifying C-2

Constraint C-2 states that: at $T_C$, the processor state is QED consistent. Further, after $T_C$, only Symbolic QED instructions can write to architectural state (test modes such as scan need to be turned off). We can specify C-2 using the below statements:

$$\uparrow(\text{SIF}_{\text{complete}}) \rightarrow \forall (i, j) \in M_r, (R_i = R_j)$$
$$\uparrow(\text{SIF}_{\text{complete}}) \rightarrow \forall (i, j) \in M_m, (\text{Mem}_i = \text{Mem}_j).$$
$$\uparrow(\text{Clock}) \rightarrow \text{if}(\text{SIF}_{\text{complete}}) \text{Test}_{\text{enable}} = 0$$

Above, $\uparrow$(signal_name) is true when signal_name transitions to high from low, and $M_r$ ($M_m$) is the set of all mapped (original, duplicate) pairs of registers (memory locations). Importantly, if there are multiple test modes, the $\text{Test}_{\text{enable}}$ signal for all of them need to be set to low. These statements are specified to the BMC tool by writing them in the form of *assume statements* (e.g., in System Verilog). Note that these constraints take the same form for both in-order and superscalar cores.

### C.3. Specifying C-3

Constraint C-3 states that: All Symbolic QED operands, must have one of the following properties: i) if the operand has already read source data then this data must match the corresponding register/memory location data at $T_C$; or ii) if the operand is waiting at $T_C$, then it is waiting on an earlier Symbolic QED instruction.

Constraint C-3(ii) is vacuously true for in-order pipelines, as an instruction only makes progress when all its operands have already read their respective data, otherwise the instruction just stalls for operand data. We use the information obtained from the Symbolic QED operand recorder (Fig. 4) to specify C-3 to the BMC tool using the below statements:

$$\uparrow(\text{SIF}_{\text{complete}}) \rightarrow \forall s \in \text{src1\_buffer} \, (s.\text{data} == \text{Reg}[s.\text{addr}])$$

$$\uparrow(\text{SIF}_{\text{complete}}) \rightarrow \forall s \in \text{src2\_buffer } (s.\text{data} == \text{Reg}[s.\text{addr}])$$
$$\uparrow(\text{SIF}_{\text{complete}}) \rightarrow \forall m \in \text{mem\_buffer } (m.\text{data} == \text{Mem}[m.\text{addr}]).$$

Above, *s.data* gives data stored for entry *s* in the buffer, while *s.addr* gives the address. *Reg* represents the architectural register array while *Mem* represents the architectural memory array. These statements are specified to the BMC tool by writing them in the form of *assume* statements (e.g., in System Verilog). For a superscalar core, specifying Constraint C-3(i) is the same as above, but we need additional information, as discussed in Sec. 3.2. to specify Constraint C-3(ii) to the BMC tool.

### C.4. Finding Counter-Examples using BMC

The QED property (see Sec. 2.2) used by BMC to find counter-examples in Symbolic QED is modified to support symbolic initial states and is given as below:

$$qed\_ready \ \& \ SIF\_complete \rightarrow \bigwedge_{a \in \{0, \ldots, \frac{N}{2}-1\}} \text{Ra} == \text{Ra}',$$

here the only change is the addition of the $SIF\_complete$ (as given in Sec. 3.2.) precondition to the QED property.